\documentclass[prl,twocolumn,showpacs,amsmath,amssymb]{revtex4}
\usepackage{graphicx}

\usepackage{dcolumn}
\usepackage{bm}

\begin{document}
\title{Avalanches in Strained Amorphous Solids: Does Inertia Destroy Critical Behavior?}
\author{K. Michael Salerno, Craig E. Maloney, and Mark O. Robbins}
\affiliation{Department of Physics and Astronomy, Johns Hopkins University, Baltimore, Maryland 21210 USA }
\affiliation{Department of Civil Engineering, Carnegie Mellon University, Pittsburgh, Pennsylvania 15213}


\date{\today}

\begin{abstract}
Simulations are used to determine the effect of inertia on athermal shear
of a two-dimensional binary Lennard-Jones glass.
In the quasistatic limit, shear occurs through a series of rapid avalanches.
The distribution of avalanches is analyzed using finite-size scaling with
thousands to millions of particles.
Inertia takes the system to a new underdamped universality class rather than
driving the system away from criticality.
Scaling exponents are determined for the
underdamped and overdamped limits and
a critical damping that separates the two regimes.
Systems are in the overdamped universality class even when most vibrational
modes are underdamped.
\end{abstract}
\maketitle 

Many slowly driven physical systems exhibit long quiescent periods punctuated
by rapid avalanches \cite{Sethna2001}.
Phenomena as diverse as earthquakes, Barkhausen noise in magnetic materials,
dislocation cascades in single crystal microcompression and fluid interface
depinning \cite{B.GUTENBERG10011944, PhysRevE.54.278, PhysRevB.52.12644, PhysRevLett.100.155502,Miguel2001,martys91,ji92} display power law avalanche statistics
in seismicity, acoustic emission, slip, stress drop or interface
advance.
These power laws reflect a non-equilibrium critical transition
at the onset of motion.

This Letter addresses a fundamental question about
the effect of inertia on such critical behavior.
Power law scaling has normally been observed in overdamped systems.
Studies of underdamped systems suggest that any inertia may drive
the system away from the critical point \cite{PhysRevA.45.665,Held1990}.
In sandpiles, the onset of motion appears to become a hysteretic first-order
transition
\cite{Jaeger1989}.
In the Burridge-Knopoff model, inertia
leads to a growing importance of
non-critical, system spanning events \cite{PhysRevLett.62.2632}.
These conclusions about the effect of inertia
seem at odds with the observation of power law scaling in earthquakes and
laboratory compression tests
\cite{B.GUTENBERG10011944,Miguel2001}, 
where seismic waves and acoustic emission
indicate that the systems are underdamped.

Here, quasistatic simulations of sheared glassy solids
are performed over a full range of damping rates.
The results reveal a rich phase diagram.
Different universality classes describe the overdamped and underdamped limits,
but both are described by critical finite-size scaling relations.
The transition between the two limits occurs at a fixed damping rate that
appears to have its own scaling behavior.
Overdamped scaling extends to surprisingly small damping rates,
where nearly all vibrational modes are underdamped.
The power law describing underdamped avalanches is close to
the Gutenberg-Richter law and an excess of large events is observed
that is similar to observations of individual fault systems.

Since we are interested in the general question of how inertia affects
critical behavior, we consider a two-dimensional binary
mixture of particles
that has been widely studied as a model amorphous system
\cite{PhysRevLett.93.016001,PhysRevLett.98.095501,PhysRevLett.104.025501}.
The particles may represent atoms, grains, bubbles, colloids  
or volume elements of a deforming fault zone.
As particle size increases, temperature becomes less relevant.
We focus on the athermal limit because it allows clear identification of
small avalanches, and because other work indicates that temperature may 
also drive systems away from criticality \cite{PhysRevE.76.036104,Lerner_thermal10}.

Particles interact via the Lennard-Jones (LJ) potential,
$ U(\bold{r}) = 4 u_0 [(a_{ij}/r)^6 - (a_{ij}/r)^{12}] $
where $r$ is the magnitude of the vector $\bold{r}$ between two particles
and the species i,j are of two types, A and B.
The particles have diameter $ a_{AA} = 5/3\ a_{BB} = a $
and $a_{AB}= 4/5\ a$.
The LJ energy and force are taken smoothly to zero at $r_c=1.5a_{ij}$
using a polynomial fit starting at 1.2$a_{ij}$ \cite{Maloney2008-JoP}.
Both particle types have mass $m$ and the number ratio
$N_A/N_B=(1+\sqrt{5})/4$.
The depth of the inter-atomic potential $u_0$ sets the energy scale of
particle interactions.
The natural unit of time
is $\tau = \sqrt{m a^2/u_0}$.

Initial states are prepared as in Ref. \cite{Maloney2008-JoP},
but, as there, the protocol has little effect on steady state shear.
After annealing, the system contains $N (\sim 10^3  - 10^6)$
particles in a square unit cell with edge $L = 27a$ to $875a$.
The density $\rho =1.38 a^{-2}$ and the pressure is near zero.
A pure shear strain is applied to the system by changing the
periodic boundaries while holding area constant.
The strain rate $|\dot{\epsilon}| < 10^{-6} \tau^{-1}$ between
avalanches was adjusted for each $L$
to ensure simulations were in the quasistatic limit where results depend
only on strain interval and not independently on time. 
When an increase in kinetic energy indicated the onset of plastic
deformation, $|\dot{\epsilon}|$ was reduced to zero to allow the
avalanche to evolve without external perturbation.
Shearing resumed after the kinetic energy dropped below 1\% of
of the background value during shear.

In order to model the athermal limit, the kinetic energy released during
avalanches must be removed by some damping mechanism.
Unless noted,
a viscous drag force was applied to each particle
$\bold{F}_{drag} = -\Gamma m \bold{v}$
where $\bold{v}$ is the non-affine velocity.
As the dissipation rate $\Gamma$ decreases,
the dynamics changes from overdamped to 
underdamped (inertial) dynamics.
We also show results for energy minimization dynamics that correspond
to $\Gamma \rightarrow \infty$.
Vibrational modes with frequency $\omega > \Gamma$ are underdamped,
and it is useful to compare $\omega$ to the root mean squared or Einstein
frequency  $\omega_E \equiv \sqrt{\left< \omega ^2 \right>} = 17 \tau^{-1}$.

We focus on the steady state achieved after plastic rearrangements
have erased memory of the initial state ($\epsilon >7$\%).
While the system is trapped in a local energy minimum, work done
by the applied strain leads to a nearly linear rise in potential
energy and shear stress $\sigma_s$.
When the minimum becomes unstable, there is a rapid avalanche
of activity that leads to a sharp drop by $E$ in energy
and by $\Delta \sigma_s$ in shear stress.
In the overdamped limit,
the system is trapped in the next local energy minimum.
When damping is reduced, inertia
can carry particles over subsequent energy
barriers to reach lower energy states.
One dramatic consequence is that the mean
energy sampled by systems 
decreases by 30\% as $\Gamma$ decreases.
Indeed, there is almost no overlap
between the ranges of energy sampled for the three damping
rates studied in detail below, $\Gamma \tau = 1$, 0.1 and 0.001.

To quantify the distributions of energy drops, we define the event rate $R(E,L$)
as the number of events of energy $E$ per unit energy and unit strain in
a system of size $L^d$ with dimension $d=2$.
A sum rule relates $R(E,L)$ to the distribution of stress drops if
$\Delta \sigma_s$ is much smaller than the mean shear stress $\left< \sigma_s \right>$
and the shear modulus $\mu$ is nearly constant \cite{PhysRevE.79.066109}.
The energy dissipated by avalanches comes from work on the system during stress increases.
We introduce an elastic energy
$S \equiv \Delta \sigma_s L^d \left< \sigma_s \right> /4\mu$
associated with a given stress change,
where
$\left< \sigma_s \right> /4\mu \approx 0.02$ for all $\Gamma$ and is independent of $L$.
In steady state, the sum over stress drops equals the sum over stress increases
and energy conservation requires:
$\int{dE E R(E,L)}=\int{dS S R(S,L)}= L^d \left< \sigma_s \right>$,
where the last quantity is the total work per unit strain.
Given the scaling exponents obtained below,
both integrals are dominated by large events
and thus $R(E,L)$ and $R(S,L)$ must exhibit the same scaling
for large avalanches.
Direct comparison of $E$ and $S$ for individual events shows
that they are not proportional for small events and we consider
the scaling of both distributions below.

Avalanche distributions are analyzed with finite-size scaling
methods \cite{privman1990finite}
that assume
the maximum size of events $\sim L^\alpha$ is limited only
by the system size $L$.
The equations are developed for $E$ but also apply to $S$.
The distribution is assumed to obey the finite-size scaling ansatz:
\begin{equation}
R(E,L) = L^{\beta}  g(E/L^{\alpha}) \ \ \ ,
\label{fss}
\end{equation}
where $g$ is an unknown scaling function.
Using the sum rule in the previous paragraph,
one finds
a scaling relation $\beta+2\alpha=d=2$ as long as the integral 
of $xg(x)$ is well defined.

Equation \ref{fss} produces power law scaling for $E << L^\alpha$ if
$g(x) \propto x^{-\tau}$ for $x << 1$.
One finds
$R(E,L) \propto L^{\gamma} E^{-\tau}$ with $\gamma \equiv \beta + \alpha \tau$.
One might expect that the probability of a small
event in a given region would be independent of system size.
This would imply a hyperscaling relation
$\gamma \equiv \beta + \alpha \tau = d$ and $\tau=2$,
given the above relation $\beta+2\alpha=d$.
As shown below, $\tau < 2$ and the number of small events rises
much less rapidly than $L^d$ in our simulations ($\gamma < d$).
Thus as $L$ increases large events suppress small
events either by changing the local configurations so small
events are less likely to occur, or by increasing the probability that
the same local configuration will produce a large avalanche.
To our knowledge, the same behavior is not observed in other systems that
display power law avalanche distributions.
For example, the probability of small events is proportional to
the size of the interface
in models of fluid invasion or domain
wall motion \cite{martys91,ji92}.

\begin{figure}[h,t,b]
\includegraphics[width=0.3\textwidth]{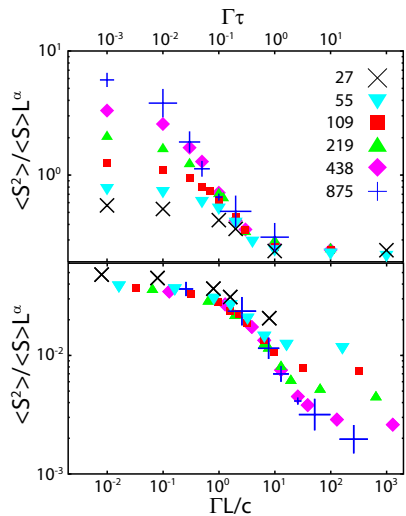}
\caption{Size of stress drops normalized
by $L^\alpha$ plotted against (a) $\Gamma \tau$ with $\alpha=0.85$
and (b) $\Gamma L/c$ with $\alpha=1.6$.
Statistical errors are smaller than the symbols for the indicated $L/a$.
}
\label{fig:Gamma}
\end{figure}

We first show how $\Gamma$ influences the 
scaling of large events with $L$.
The mean size $\left< S \right>$ is not well-defined because
of the diverging number of small events for $\tau \geq 1$.
Instead we show the ratio $\left< S^{2} \right>/\left< S \right> L^\alpha$,
which gives similar event sizes and scaling as other moment ratios.
As shown in Fig. \ref{fig:Gamma}(a), $\alpha=0.85$ collapses
results for all $L$ onto a universal curve
for $\Gamma \tau > 0.1$.
Results for $\Gamma \tau \geq 1$ and energy minimization
are statistically indistinguishable.
As $\Gamma \tau $ decreases from 1 to 0.1,
the mean avalanche size increases in the same way for all $L$.
Since $\omega_E \tau =17$,
systems remain in the overdamped universality class even when
almost all vibrational modes are underdamped.
The key factor is not whether modes are overdamped but whether
inertia can carry the system over the next energy barrier in the
energy landscape.
A small $\Gamma$ can trap the system in the nearest minimum if
the difference in successive energy
barriers is small and/or the path in phase space to the
next barrier is complicated.

As $\Gamma \tau$ decreases below 0.1 in Fig. \ref{fig:Gamma}(a), results for
different $L$ separate.
As shown in Fig. \ref{fig:Gamma}(b),
results in this underdamped limit can be collapsed using
$\alpha=1.6$ and scaling $\Gamma$ by the time $L/c$ for sound propagation
across the system at shear velocity $c= 3.4 a/\tau$.
Data for $L>27a$ follow a common scaling curve until 
$\Gamma \tau$ exceeds 0.1, and the critical behavior changes to overdamped.
For $\Gamma L/c <<1$, all sound waves in the system are underdamped and
all results tend to the same value of
$\left< S^2 \right>/\left< S \right>L^\alpha$.
As $\Gamma L/c $ increases past unity, the longest wavelength modes
begin to be overdamped.
The scaling exponent does not change, but
$\left< S^2 \right>/\left< S \right> L^\alpha$ decreases.
This is consistent with avalanches being cutoff by the wavelength
of the largest underdamped mode rather than the system size.
For the damping used here, this wavelength scales like 1/$\Gamma$ 
and any finite damping takes the system away from the underdamped
critical point.
However, long wavelength modes are always underdamped as $L \rightarrow 0$
in disordered solids because dissipation mechanisms must be
Galilean-invariant and only damp relative velocities \cite{landaulif}.
Weakly damped
simulations with two different Galilean-invariant thermostats
\cite{Maloney2008-JoP,DPD_Hoogerbrugge92} gave
statistically identical results to those shown in Figs. 1-3
for $\Gamma L/c < 0.1$.

\begin{figure}[h,t,b]
\includegraphics[width=0.4\textwidth]{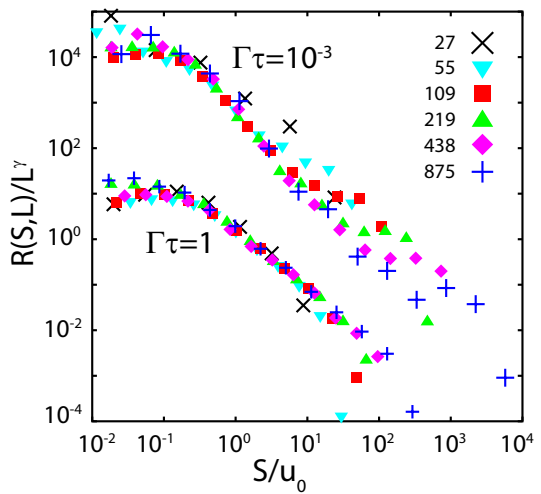}
\caption{Scaling of $R(S,L)$ with $S$ for overdamped $\Gamma=1$ systems
with $\gamma=1.3$ (lower curves)
and underdamped $\Gamma=0.001$ systems with $\gamma=1.2$ (upper curves).
Underdamped curves are multiplied by 100 to prevent overlap,
symbols indicate $L/a$ and symbol size is larger than statistical errors.
}
\label{fig:unscaled}
\end{figure}

We now examine the avalanche scaling in more detail in the
overdamped ($\Gamma\tau=1$) and underdamped ($\Gamma\tau=0.001$)
limits and at the crossover between them ($\Gamma\tau=0.1$).
Figure \ref{fig:unscaled} shows $R(S,L)/L^{\gamma }$ vs. $S$
in an overdamped system (lower curves).
Power law scaling is observed from $S \sim 0.2 u_0$ to a cutoff that
increases with $L$.
Data in this scaling regime are collapsed with
$\gamma = 1.3 \pm 0.1$.
As noted above, $\gamma < d$ implies that the number of events 
per unit area at a given $S$ is strongly suppressed as $L$ increases.

\begin{figure}[h,t,b]
\includegraphics[width=0.4\textwidth]{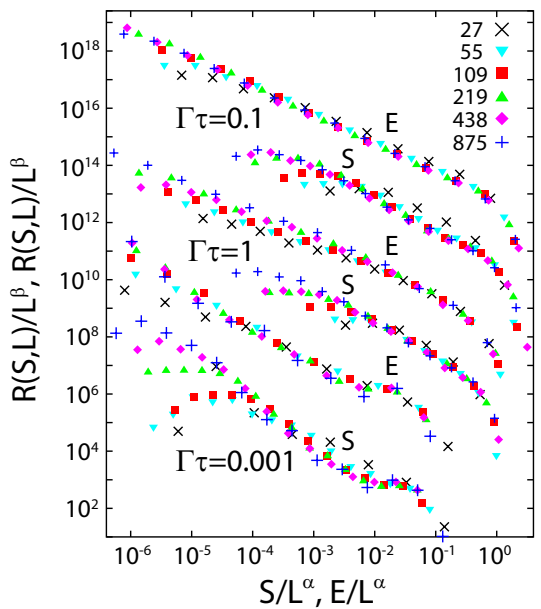}
\caption{(a) Finite-size scaling collapse of
$R(S,L)$ and $R(E,L)$ for overdamped systems ($\Gamma \tau=1$),
underdamped systems ($\Gamma \tau=0.001$)
and critically damped systems ($\Gamma \tau =0.1$)
using the exponents in Table \ref{tab:exponents}.
Statistical errors are smaller than the symbols and successive
curves are shifted up by 2 or 3 decades to prevent overlap.
}
\label{fig:fss}
\end{figure}

Figure \ref{fig:fss} shows finite-size scaling collapses of overdamped
data for both $E$ and $S$ (middle curves).
The scaling factor in $S$ was chosen so that the two quantities are comparable
for large events, 
but the correlation between $E$ and $S$ breaks down below $0.2u_0$.
Both quantities are well described by common scaling exponents Table (\ref{tab:exponents}) from
$0.2 u_0$ up to the largest event sizes.

At lower energies, $R(S,L)$ saturates while $R(E,L)$ follows a different
power law that changes slightly with $L$.
It is easy to confuse this power law with critical scaling
if one only has results for $R(E,L)$ at small $L$.
The noncritical power law dominates $R(E,L)$ for the smaller system sizes
$L<109a$ used in previous studies.
This explains deviations in the reported values of $\alpha$
\cite{PhysRevLett.93.016001,PhysRevE.79.066109} and
why some papers concluded there was no
critical behavior \cite{Tewari1999,Hatano2009}.
Dahmen has recently suggested that $\tau$ has
a mean-field value of 1.5 for all $d$ in overdamped systems 
\cite{Dahmen2011}.
Our result of 1.2 is lower, but substantially higher
than the values of $\tau<1$ that would be inferred from the noncritical
power law region in $R(E,L)$
\cite{PhysRevLett.93.016001,Tewari1999}.


Figure \ref{fig:unscaled} also shows $R(S,L)/L^\gamma$ for underdamped
systems (upper curves).
As expected, inertia leads to much larger avalanches.
Results in
the scaling regime for $S > 0.2 u_0$ collapse with $\gamma=1.2$.
The distributions all show a power law decay followed by
a plateau that moves to larger $S$ as $L$ increases.
While this is different from the sharp cutoff in the underdamped case, the
form of the scaling function $g(x)$ in Eq. \ref{fss} need not be simple.
Figure \ref{fig:fss} shows finite-size scaling collapses
for $S$ and $E$ with the same scaling exponents (lower curves).
In both cases, results for large events from different $L$ collapse
onto a universal curve.
There is a plateau over a fixed range of a little under a decade
followed by a very rapid decrease.

These results clearly imply that inertia does not destroy critical
behavior, but does lead to a different universality class.
Results for Galilean invariant thermostats with weak damping show
the same scaling behavior.
While our system lacks the complexity found in earthquake faults,
it is interesting to note that $\tau$ is close to the value
of $\sim 1.6$ for the Gutenberg-Richter law \cite{scholz2002mechanics}.
In addition, the distribution of earthquakes for a given fault system
typically has an excess of large events that is similar to the plateau
seen in Fig. \ref{fig:fss} \cite{scholz2002mechanics}.

The final example we consider is the intermediate case of
$\Gamma\tau =0.1$ that seems to represent a crossover between overdamped
and underdamped scaling in Fig. \ref{fig:Gamma}.
Figure \ref{fig:fss} shows a finite-size scaling
collapse of $R(E,L)$ and $R(S,L)$ (top curves).
The results follow a power law with $\tau=1$ over 6 decades or more.
Similar scaling was found for intermediate damping with Galilean invariant
thermostats,
for different interaction potentials, for simple shear, and in preliminary
studies of 3D systems.
This suggests that $\Gamma \tau =0.1$ represents
a multicritical point separating
regions that flow to underdamped
and overdamped fixed points.


\begin{table}
\begin{tabular}{| c | c | c | c | c |}
\hline
$\Gamma$ & $\tau$ & $\alpha$ & $\beta$ & $\gamma $ \cr
\hline
 1.0 & $1.2 \pm 0.05 $ & $0.85 \pm 0.05 $ & $0.3 \pm 0.05 $ & $1.3 \pm 0.05 $\\
 0.1 & $1.0 \pm 0.05 $ & $0.85 \pm 0.05 $ & $0.4 \pm 0.05 $ & $1.2 \pm 0.05 $ \\
 0.001 & $1.5 \pm 0.1 $ & $1.6 \pm 0.1 $ & $-1.2 \pm 0.1 $ & $1.2 \pm 0.1 $ \\
\hline
  \end{tabular}
\caption{
Scaling exponents determined for overdamped ($\Gamma=1$)
and underdamped ($\Gamma=0.001$) limits and at the crossover
between them $\Gamma=0.1$.
Quoted values satisfy the scaling relations
$\beta + 2\alpha=d$ and $\gamma=\beta+\alpha \tau$ and
errorbars are estimated from the quality of finite-size scaling
collapses for $E$ and $S$ using other $\Gamma$ and moments.
}
\label{tab:exponents}
\end{table}

In conclusion, introducing inertia does not destroy
critical scaling of avalanches in quasistatic shear of disordered solids.
Systems continue to be in the overdamped universality class even when
most vibrational modes are underdamped.
Only a small amount of damping is needed to prevent inertia from carrying
systems over sequential energy barriers,
implying that the difference between energy barriers is small or
the path between them complex.
Below a critical damping rate a new universality class corresponding
to the underdamped limit is identified.
The exponent describing avalanches is close to the Gutenberg-Richter law and
the finite-size scaling function has an unusual form with a plateau
before the cutoff at large events.
Different scaling exponents are observed at the critical damping rate,
indicating that it is a multicritical point.

The scaling exponents in all three regimes (Table \ref{tab:exponents})
satisfy the scaling relations $\beta +2\alpha=d$ and
$\gamma=\beta+\alpha \tau$.
The hyperscaling relation $\gamma =d$ is violated in all cases.
The number of avalanches at a given energy rises less rapidly
than system size ($\gamma <d$), indicating that small events
are suppressed by the larger events in bigger systems.
Exponents obtained from finite-size scaling of
the distribution of energy and stress drops are
consistent.
However, there is a long power law tail in $R(E,L)$ at small $E$ with a size dependent
exponent and system-size independent cutoff.
This tail appears to have dominated previous determinations
of $\alpha$ \cite{PhysRevLett.93.016001,PhysRevE.79.066109}
and $\tau$ \cite{Tewari1999,PhysRevE.74.016118} using
smaller systems.

\begin{acknowledgments}
We thank Karin Dahmen for useful discussions.
This work was supported by the National Science Foundation (NSF)
under grants DMR-10046442, CMMI-0923018, and OCI-108849.
\end{acknowledgments}

\bibliography{newav9}{}

\bibliographystyle{apsrev}

\end{document}